\documentclass[aps,prl,showpacs,twocolumn,superscriptaddress,preprintnumbers]{revtex4-2}

\RequirePackage{subfigure}
\usepackage{lipsum}
\usepackage{graphicx}
\usepackage{color}
\usepackage{enumerate}
\usepackage{amsmath}
\usepackage{amsthm} %
\usepackage{bm}
\usepackage{diagbox}
\usepackage{amssymb}
\usepackage{dsfont}
\usepackage{stmaryrd}
\usepackage{multirow}
\usepackage{bbm}
\usepackage[normalem]{ulem}
\usepackage{float}
\usepackage{longtable,booktabs}
\usepackage[dvipsnames,table,xcdraw]{xcolor}
\usepackage[normalem]{ulem}
\usepackage{physics}
\usepackage{url}
\usepackage{soul} 
\usepackage[colorinlistoftodos]{todonotes}
\usepackage{verbatim}
\usepackage{graphicx}
\usepackage{subfigure}
\graphicspath{{figures/}}
\usepackage{appendix}
\definecolor{darkblue}{rgb}{0.1,0.2,0.6} \definecolor{darkred}{rgb}{0.8,0.1,0.2}
\usepackage[colorlinks,citecolor=darkblue,linkcolor=darkred,urlcolor=darkblue]{hyperref}
\usepackage[all]{hypcap} 
\usepackage[draft]{pdfcomment}

\newcommand\id{\ensuremath{\mathbbm{1}}} 

\hypersetup{
	bookmarks=false,         %
	unicode=false,          %
	pdftoolbar=false,        %
	pdfmenubar=true,        %
	pdffitwindow=false,     %
	pdfstartview={FitH},    %
	pdftitle={},    %
	pdfauthor={Authors},     %
	pdfsubject={},   %
	pdfcreator={},   %
	pdfproducer={}, %
	pdfnewwindow=true,      %
	colorlinks=true,       %
	linkcolor=black,          %
	citecolor=blue,        %
	filecolor=magenta,      %
	urlcolor=blue           %
}

\usepackage{xcolor}

\usepackage{tikz}
\usepackage{tikz-cd}
\usetikzlibrary{arrows}
\usetikzlibrary{intersections}
\usetikzlibrary{shapes.geometric}
\usetikzlibrary{decorations.pathmorphing, patterns,shapes}
\usetikzlibrary{decorations.markings}

\tikzset{
	partial ellipse/.style args={#1:#2:#3}{
		insert path={+ (#1:#3) arc (#1:#2:#3)}
	}
}

\tikzset{
	mid arrow/.style={postaction={decorate,decoration={
				markings,
				mark=at position .575 with {\arrow[#1]{stealth}}
	}}},
	near arrow/.style={postaction={decorate,decoration={
				markings,
				mark=at position .275 with {\arrow[#1]{stealth}}
	}}},
	far arrow/.style={postaction={decorate,decoration={
				markings,
				mark=at position .800 with {\arrow[#1]{stealth}}
	}}},
}

\tikzset{ 
    table/.style={
        matrix of nodes,
        row sep=-\pgflinewidth,
        column sep=-\pgflinewidth,
        nodes={
            rectangle,
            draw=black,
            align=center
        },
        minimum height=1.5em,
        text depth=0.5ex,
        text height=2ex,
        nodes in empty cells,
        column 1/.style={
            nodes={text width=2em,font=\bfseries}
        },
        }
    }

\newcommand{\calN}{\mathcal{N}}
\newcommand{\calO}{\mathcal{O}}

\newcommand{\ri}{\mathrm{i}}

\begin{document}
\title{Optimized trajectory unraveling for classical simulation of noisy quantum dynamics}

\author{Zhuo Chen}
\affiliation{Center for Theoretical Physics, Massachusetts Institute of Technology, Cambridge, MA 02139, USA}
\affiliation{The NSF AI Institute for Artificial Intelligence and Fundamental Interactions, Cambridge, MA 02139, USA}
\author{Yimu Bao}
\affiliation{Department of Physics, University of California, Berkeley, California 94720, USA}
\author{Soonwon Choi}
\affiliation{Center for Theoretical Physics, Massachusetts Institute of Technology, Cambridge, MA 02139, USA}


\begin{abstract}
The dynamics of open quantum systems can be simulated by unraveling it into an ensemble of pure state  trajectories undergoing non-unitary monitored evolution, which has recently been shown to undergo measurement-induced entanglement phase transition.
Here, we show that, for an arbitrary decoherence channel, one can optimize the unraveling scheme to lower the threshold for entanglement phase transition, thereby enabling efficient classical simulation of the open dynamics for a broader range of decoherence rates.
Taking noisy random unitary circuits as a paradigmatic example, we analytically derive the optimum unraveling basis that on average minimizes the threshold.
Moreover, we present a heuristic algorithm that adaptively optimizes the unraveling basis for given noise channels, also significantly extending the simulatable regime.
When applied to noisy Hamiltonian dynamics, the heuristic approach indeed extends the regime of efficient classical simulation based on  matrix product states beyond conventional quantum trajectory methods.
Finally, we assess the possibility of using a quasi-local unraveling, which involves multiple qubits and time steps, to efficiently simulate open systems with an arbitrarily small but finite decoherence rate.
\end{abstract}

\maketitle

\textit{Introduction.---}
While an ideal quantum system evolves under unitary evolution according to the Schr\"{o}dinger's equation,
realistic quantum systems are inevitably subject to environmental noise and undergo open system dynamics, often well modeled by the Lindblad equation~\cite{breuer2002theory}.
Efficient classical simulation of such dynamics not only enables theoretical understanding of open quantum systems but is also pivotal in advancing quantum simulation experiments~\cite{arute2019quantum,Wu_2021,google2023} and benchmarking near-term quantum devices~\cite{choi2023preparing,cotler2023emergent}.

Stochastic wavefunction method is a widely-adopted approach to simulate open system dynamics~\cite{PhysRevA.45.4879,dalibard1992wave,PhysRevA.46.4363,plenio1998quantum,carmichael2009open}. 
In this method, instead of simulating the evolution of the mixed density matrix describing the open system, one \emph{unravels} the mixed state into an ensemble of pure quantum trajectories whose statistical average emulates the evolution of the mixed state.
Interestingly, each trajectory in the ensemble follows non-unitary evolution that can be understood as monitored quantum dynamics, wherein entanglement can undergo a measurement-induced phase transition from a volume- to an area-law scaling when the noise or decoherence rate increases~\cite{li2018quantum,skinner2019measurement,li2019measurement,choi2020quantum,szyniszewski2019entanglement,gullans2020dynamical}.
Since the simulation cost grows exponentially with entanglement entropy and, in particular, the area-law scaling of entanglement implies efficient representation of quantum states in one dimension using matrix product states (MPS)~\cite{PhysRevLett.69.2863, PhysRevLett.91.147902, Vidal_2004, SCHOLLWOCK201196},  
this phenomenon indicates the presence of a critical decoherence rate  above which efficient classical simulation is feasible in one dimension \cite{azad_hallam_morley_green_2023}.

Importantly, the same open system dynamics can be unraveled into infinitely many equivalent trajectory ensembles, and the entanglement entropy within trajectories strongly depends on the unraveling scheme. 
In a pioneering work, Ref.~\cite{vovk2022entanglement} focused on two classes of unraveling schemes and designed an entanglement-optimized algorithm that minimizes the entanglement growth at each time step.
Independently, Ref.~\cite{kolodrubetz2023optimality} considered random unitary circuits (RUC) with dephasing noise and focused on four empirically chosen unraveling schemes, and found that two of them result in a lower critical decoherence rate compared to the conventional unraveling using projective measurements.
Yet, it remains open, for generic quantum evolution subject to a generic type of decoherence, whether one can systematically obtain an unraveling scheme that minimizes the critical decoherence rate and extends the regime of efficient simulation.

\begin{figure}
    \centering
    \includegraphics[width=\linewidth]{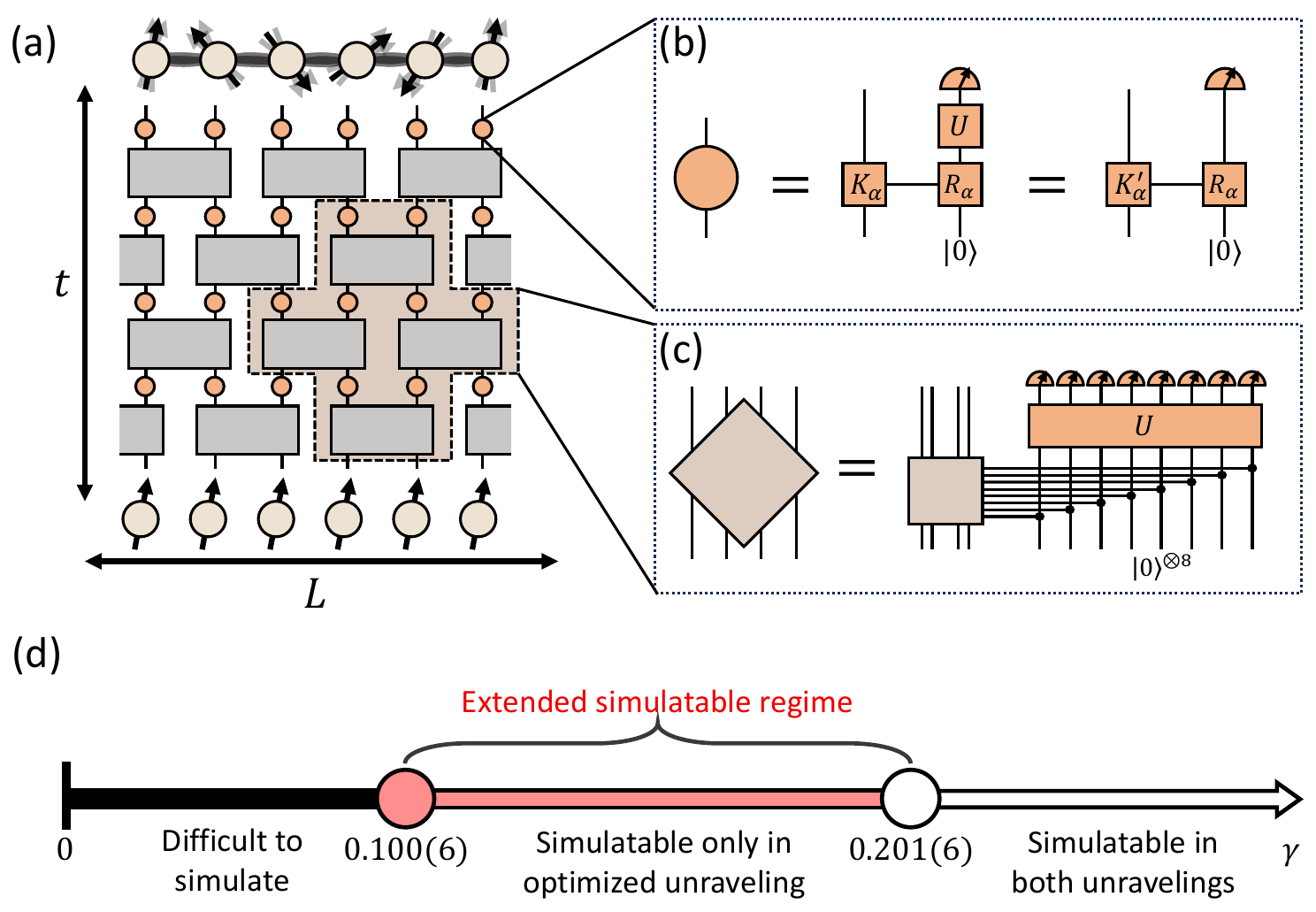}
    \caption{(a) Quantum evolution subject to generic local decoherence. The time evolution is digitized into a quantum circuit with local decoherence channels. (b) The decoherence channel can be expressed as a coupling to an ancilla qudit and measuring the ancilla qudit, where the coupling gate is a unitary operator $\sum_\alpha K_\alpha \otimes R_\alpha$ that maps $\ket{\psi}\otimes\ket{0}$ to $\sum_\alpha K_\alpha \ket{\psi} \otimes \ket{\alpha}$. Equivalent unravelings can differ by a unitary gate $U$ on the ancilla qudit, which results in an equivalent set of Kraus operators $K'_\alpha = \sum_\beta U_{\alpha\beta} K_\beta$. (c) Multiple unitary gates and decoherence channels can be combined into a two-qudit quantum channel, which can be purified by applying a joint unitary gate on the ancilla and system qudits.
    (d) Phase diagram for simulating mixed-field Ising model with dephasing noise using trajectory unraveling. The unraveling in computational basis yields a critical error rate of $\gamma_c=0.201(6)$, whereas the optimized unraveling reduces the critical rate down to $\gamma_c=0.100(6)$.
    }
    \label{fig:1}
\end{figure}

In this paper, we show that, for arbitrary types of decoherence, one can obtain an unraveling scheme that optimizes the critical decoherence rate and therefore can extend the regime of efficient classical simulation.
We consider the most general form of unraveling schemes that involve a minimum number of Kraus operators for each decoherence channel.
As a paradigmatic example, we first focus on noisy random unitary circuits [Fig.~\ref{fig:1}(a)] and obtain an analytic expression for an optimized unraveling basis that minimizes the critical threshold on average based on an effective spin model. 
Subsequently, we propose a heuristic algorithm that adaptively optimizes the unraveling basis for a given arbitrary quantum channel to maximally disentangle individual noisy qubits from the rest in each simulation step.
Both methods significantly extend the regime of classical simulation.
We then apply our heuristic algorithm to simulate noisy Hamiltonian dynamics and demonstrate how classical simulation, utilizing MPS, becomes viable for low decoherence rates that were previously not feasible.
We further discuss the possibility to extend the efficient simulation regime to any small but finite decoherence rate based on a quasi-local unraveling scheme, where multiple qubits and time steps are jointly unraveled [Fig.~\ref{fig:1}(c)].

We note that alternative methods for simulating noisy evolution have been proposed based on matrix product operator (MPO) representation of the density matrix~\cite{verstraete2004matrix,zwolak2004mixed,noh2020efficient,white2018quantum}. 
However, these methods suffer from issues such as failure to maintain the positivity of the density matrix, resulting in negative eigenvalues, and lack of rigorous error bounds for quantum state fidelities over the dynamics.
In the Supplementary Material~\cite{SOM}, we provide comparisons of the proposed method and MPO-based method in practical settings.
Besides, Ref.~\cite{aharonov2022polynomial} showed the existence of an efficient classical algorithm for noisy random circuits with any finite decoherence rate, but the algorithm is difficult to implement in practice.

\textit{Unraveling open system dynamics.---} 
The dynamics of a one-dimensional system in a Markovian noisy environment is generally described by the Lindblad equation~\cite{breuer2002theory}.
Here, we consider a discretized evolution, in which the dissipative part of the Lindblad equation acts on the system as quantum channels~\cite{SOM} [see Fig.~\ref{fig:1}(a)].
We focus on the unraveling of these quantum channels that decohere the underlying unitary evolution.

A generic quantum channel can be always decomposed in terms of the Kraus operators $K_\alpha$
\begin{align}
    \calN[\rho] = \sum_{\alpha = 0}^{N-1} K_\alpha \rho K_\alpha^\dagger,
\end{align}
where $K_\alpha$ satisfies $\sum_{\alpha = 0}^{N-1} K^\dagger_\alpha K_\alpha = \id$.
Such a decoherence channel can be formulated as unitary coupling $U_{QM}$ between the system $Q$ and an ancilla qudit $M$; tracing out the ancilla qudit reproduces the quantum channel~\cite{preskill1998lecture}, i.e.
\begin{align}
    \calN[\rho] = \tr_M \left( U_{QM} \left(\rho \otimes \ket{0}\bra{0}\right) U_{QM}^\dagger \right),
\end{align}
where $U_{QM}$ is a suitably chosen unitary operator that maps $\ket{\psi}\otimes\ket{0}$ to $\sum_\alpha K_\alpha \ket{\psi} \otimes \ket{\alpha}$, and the ancilla qudit is of dimension $N$ same as the number of Kraus operators.

Tracing the ancilla qudit is equivalent to summing over an ensemble of trajectories generated by measuring the ancilla qudit in a computational basis.
The probability of the measurement outcome $\alpha$ is determined by the Born rule.
In each trajectory associated with the outcome $\alpha$, a Kraus operator $K_\alpha$ acts on the system. 
Such an ensemble of trajectories averages to the mixed density matrix $\calN[\rho]$ and therefore serves as an unraveling of the channel.

Crucially, the quantum channel can be unraveled in infinitely many equivalent ways.
Equivalent schemes can be obtained by measuring the ancilla qudit in a rotated basis and thus are related by a unitary transformation $U$ as in Fig.~\ref{fig:1}(b)~\footnote{In general, $U$ can be isometry. For simplicity, we only optimize over unitary transformations on the ancilla qudit.}.
Alternatively, the unitary rotation $U$ on the ancilla qudit can be viewed as decomposing $\calN$ in terms of a different set of Kraus operators related to $\{K_\alpha\}$ by $K'_\alpha = \sum_\beta U_{\alpha\beta} K_\beta$.
In the rest of the paper, we take either perspective on relating equivalent unraveling schemes, whichever is convenient.

The challenge for efficient classical simulation lies in the entanglement entropy within each trajectory.
To illustrate this, we consider the evolution subject to local dephasing noise of rate $p$, described by $\calN_{\phi,i}[\rho] = (1-p/2)\rho + (p/2) Z_i \rho Z_i$.
The channel can be unraveled into probabilistic projective measurement in the Pauli-Z basis with probability $p$.
It has been shown that such a unitary evolution interspersed by measurements exhibits a measurement-induced transition in the half-chain entanglement entropy from a volume- to an area-law scaling~\cite{li2018quantum,skinner2019measurement,li2019measurement}.
This indicates the classical simulation of trajectory dynamics is efficient only above the critical decoherence rate $p_c$.

However, $p_c$ depends on the unraveling basis.
For example, the dephasing channel $\calN_{\phi,i}$ can be also unraveled into trajectories in which a $Z_i$ gate is applied with probability $p/2$.
In this unraveling, every trajectory undergoes purely unitary evolution and is generally always in the volume-law phase, making it difficult to simulate classically.
Our goal is to find an optimized unraveling basis that minimizes $p_c$ and extends the regime of efficient simulation.

\textit{Optimized unraveling for noisy random unitary circuits.---}
We first consider the trajectory unraveling of noisy random unitary circuits (RUC) operating on a one-dimensional chain of qudits with local Hilbert space dimension $q$.
The circuit involves two-qudit Haar random unitary gates arranged in a brick-layer structure and local decoherence channels applied to every single qudit after each layer as shown in Fig.~\ref{fig:1}(a).

\begin{figure}
    \centering
    \includegraphics[width=\linewidth]{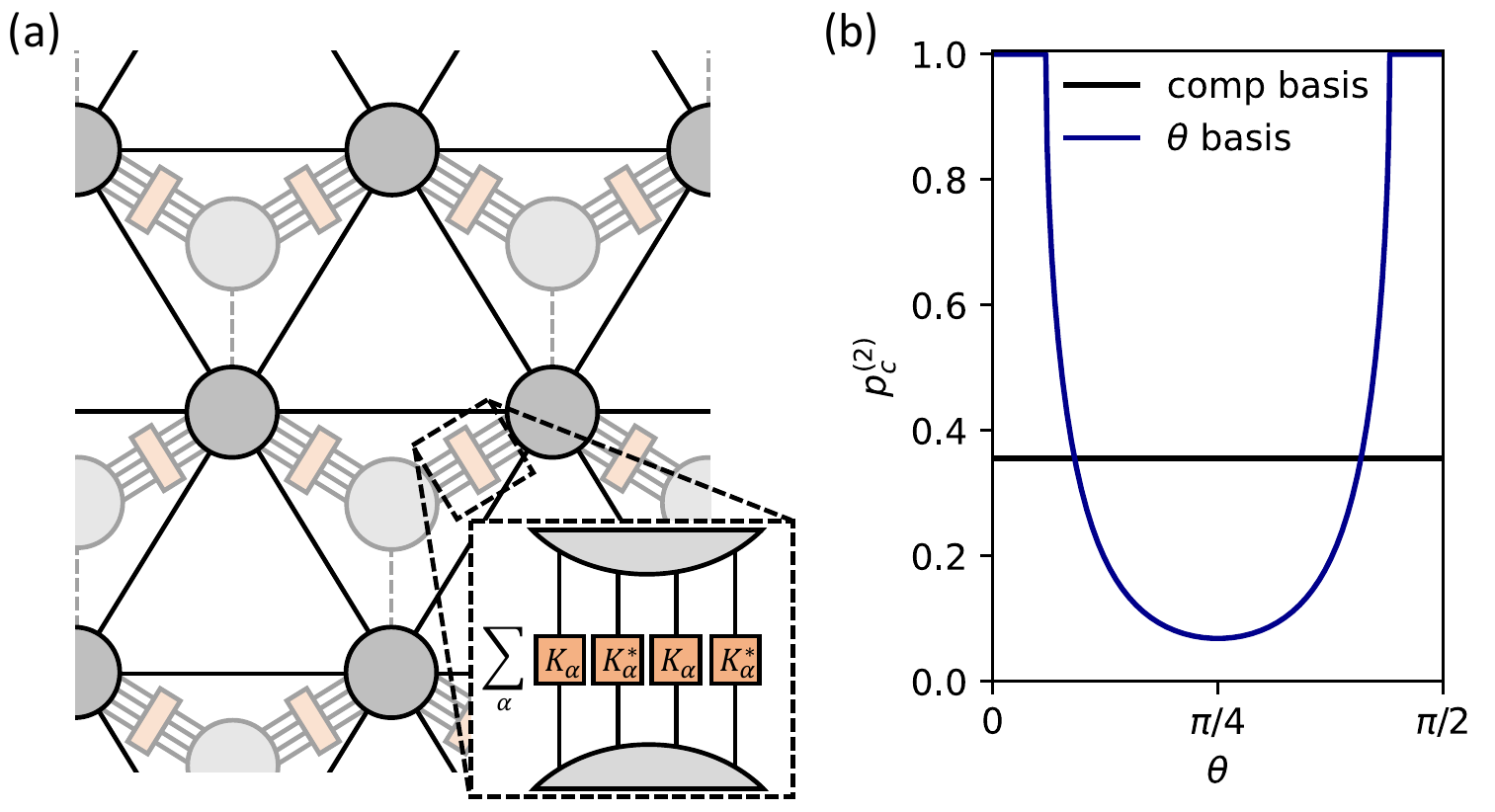}
    \caption{(a) Effective Ising spin model on a triangular lattice for noisy random unitary circuits (RUC). The Ising couplings between spins on the same downward-facing triangle depend on the unraveling scheme, i.e. the Kraus operators $K_\alpha$. (b) Critical point $p_c^{(2)}$ in the spin model for various unraveling schemes of the RUC with dephasing noise. The black and blue lines represent $p_c^{(2)}$ in the conventional unraveling based on projective measurements, and general unraveling schemes involving two Kraus operators, respectively. $p_c^{(2)}$ in the general unraveling scheme as a function of $\theta$ for fixed $\phi = \pi/4$ is presented. The lowest $p_c^{(2)}$ is obtained at $\theta=\phi=\pi/4$.
    }
    \label{fig:2}
\end{figure}

\begin{figure}
    \centering
    \includegraphics[width=\linewidth]{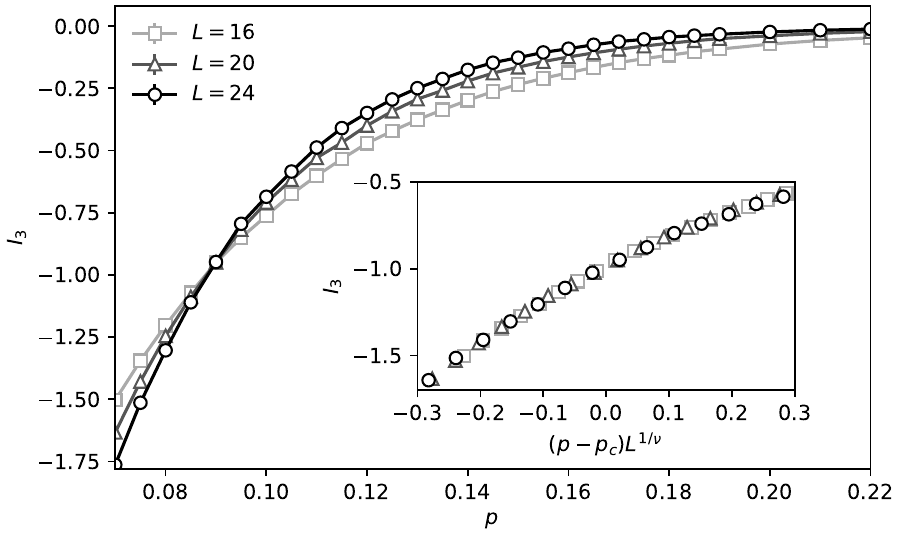}
    \caption{Tripartitie mutual information $I_3$ as a function of the decoherence rate $p$ in the spin-model optimized unraveling for RUC. The results are obtained from exact simulation up to system size $L = 24$. (Inset) Finite-size scaling collapse determines the critical decoherence rate $p_c=0.089(3)$. For comparisons, conventional unraveling based on projective measurements in computational basis yields $p_c=0.168(3)$, and the heuristically optimized unraveling basis $p_c=0.085(2)$. The results are averaged over 400 quantum trajectories.
    }
    \label{fig:3}
\end{figure}

The entanglement entropy in the trajectories of noisy Haar random circuits has an analytic albeit qualitative description in terms of the domain wall free energy in a two-dimensional classical Ising spin model on a triangular lattice [Fig.~\ref{fig:2} (a)]~\cite{bao2020theory,jian2020measurement}.
The spin model exhibits a ferromagnetic transition when tuning the decoherence rate, which is detected by the domain wall free energy and corresponds to the transition in the entanglement entropy.
The couplings in the spin model are between the neighboring spins in the triangular lattice and depend on both the decoherence rate and unraveling scheme.
Remarkably, in the case that every channel is unraveled in the same basis, the spin model is translationally invariant, and its critical point $p_c^{(2)}$ is exactly solvable~\cite{eggarter1975triangular,tanaka1978triangular}.
Thus, we can determine an optimized unraveling basis for the RUC that exactly minimizes the analytically determined critical decoherence rate $p_c^{(2)}$.
We note that the spin model only describes the quasi-entropy, and its critical point $p_c^{(2)}$ empirically approximates $p_c$ in the true von Neumann entropy.
Yet, the unraveling basis that minimizes $p_c^{(2)}$ still greatly reduces the $p_c$ for von Neumann entropy, which will be numerically verified in Fig.~\ref{fig:3}.

Specifically, for a given unraveling scheme, i.e. a set of Kraus operators $\{K_\alpha\}$ depending on the decoherence rate $p$, the critical point $p_c^{(2)}$ in the spin model can be determined from~\cite{bao2020theory}
\begin{align}
    \left(\frac{u_2}{u_1}\right)^2 - 2\frac{q^2-1}{q^2+1}\left(\frac{u_2}{u_1}\right) - 1 = 0,
\end{align}
where
\begin{align}
    u_2 &= \sum_{\alpha = 0}^{N-1} \left(\tr K_\alpha^\dagger K_\alpha\right)^2,\,\,
    u_1 = \sum_{\alpha = 0}^{N-1} \tr K_\alpha^\dagger K_\alpha K_\alpha^\dagger K_\alpha.
\end{align}
We optimize the critical threshold $p_c^{(2)}$ in the spin model over equivalent sets of Kraus operators related by unitary transformations $U$.
For the dephasing channel, it turns out $p_c^{(2)}$ depends on two parameters $\theta$ and $\phi$~\cite{SOM}, and the optimum is found at $\theta=\phi=\pi/4$ with $p_c^{(2)} = 0.0685$, which is significantly lower than $p_c^{(2)} = 0.3558$ in the conventional unraveling based on projective measurements~\cite{bao2020theory}.
The resulting Kraus operators are $K_{0,1} = (\pm\sqrt{1-p/2}\id + \sqrt{p/2}Z)/\sqrt{2}$, which coincides with those in one unraveling scheme studied in Ref.~\cite{kolodrubetz2023optimality} up to a sign difference.
Our method can be applied to an arbitrary type of decoherence channel.
For depolarization and amplitude damping noise, we obtain the spin-model optimized basis with corresponding $p_c^{(2)}$ in the Supplementary Material~\cite{SOM}. 

So far, we only consider optimizing the critical decoherence rate $p_c^{(2)}$ in the spin model.
However, the entanglement within the simulatable regime, i.e. the area-law phase, may not be optimally minimized.
With this analytic tool, it is worth exploring whether one can minimize the quasi-entropy within the area-law phase to further reduce the computational cost of classical simulation.

Exactly solving the spin model can determine an unraveling basis that is optimized on average for all random circuit realizations and trajectories.
However, in practice, the optimal basis for each decoherence channel depends on the specific quantum state it applies to and therefore can vary among different trajectories.
Here, we propose a heuristic algorithm that optimizes the unraveling for each individual decoherence channel.
The most straightforward idea is to find the unraveling basis that results in the minimum entanglement in the system.
However, such an optimization process is computationally expensive, and we instead search for the basis that maximally and locally disentangles the noisy qubit from the rest of the system.
Specifically, we first compute the reduced density matrix $\rho_i$ of the noisy qubit.
Then, we optimize over equivalent Kraus decompositions of the decoherence channel such that the average entanglement between the noisy qubit and the rest is minimized.

To demonstrate the optimized unraveling schemes indeed lowers $p_c$, we perform an exact numerical simulation of Haar random unitary circuits subject to dephasing noise operating on a chain of qubits ($q = 2$) up to system size $L = 24$ with periodic boundary condition.
We compute the tripartite mutual information $I_3=S_A+S_B+S_C-S_{AB}-S_{AC}-S_{BC}+S_{ABC}$ to determine the critical point $p_c$, where $A$, $B$, and $C$ each represents a quarter of the system.
Such a quantity is expected to change sharply from a volume-law scaling $I_3 = \mathcal{O}(L)$ to an area-law scaling $I_3 = \mathcal{O}(1)$ across the phase transition~\cite{zabalo2020critical}.
In Fig.~\ref{fig:3}, we perform the finite-size scaling using the ansatz $I_3=\mathcal{F}((p-p_c) L^{1/\nu})$ to extract $p_c$.
Compared to the conventional unraveling based on projective measurements with $p_c=0.168(3)$~\cite{SOM,zabalo2020critical}, the optimized unravelings obtained from the spin model and from the heuristic algorithm yield significantly lower critical decoherence rates $p_c=0.089(3)$ and $p_c=0.085(2)$, respectively.

\textit{Optimized unraveling for Hamiltonian dynamics with dephasing noise.---} 
The optimized unraveling scheme can also extend the regime of efficient simulation for noisy Hamiltonian dynamics.
Here, we study the one-dimensional mixed-field Ising model (MFIM) under dephasing noise, which is governed by the Lindblad equation $\dot \rho = -\ri \comm{H}{\rho} + \gamma \sum_i  (Z_i \rho Z_i^\dagger - \frac{1}{2}\{Z_i^\dagger Z_i,\rho\})$.
We consider $H = -\sum_{\langle i, j \rangle} Z_i Z_j + 1.05 \sum_i X_i - 0.5 \sum_i Z_i$, which is far from any integrable system and is challenging to simulate in the absence of decoherence~\cite{banuls2011strong}. 
We remark that the trotterized evolution of the Lindblad equation can be represented in the brick-layer structure in Fig.~\ref{fig:1}(a), where the quantum channels are $\calN_{\phi,i}$ of dephasing rate $p=2\gamma dt$. 

We first apply our heuristic algorithm to exactly simulate the noisy dynamics.
The finite size scaling analysis yields the critical decoherence rate $\gamma_c=0.100(6)$, which is significantly lower than $\gamma_c=0.201(6)$ obtained in the conventional unraveling~\cite{SOM}.

\begin{figure}
    \centering
    \includegraphics[width=\linewidth]{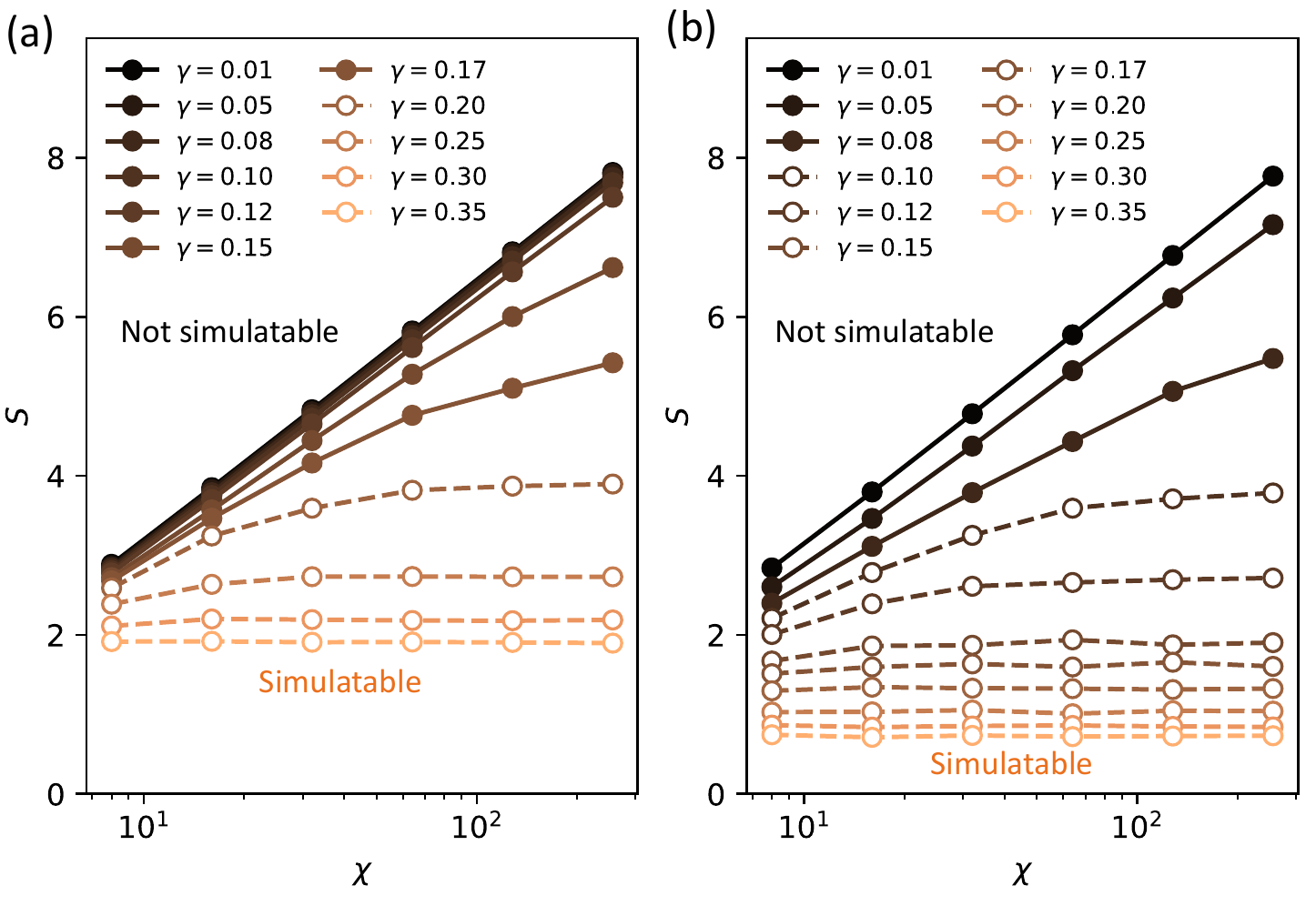}
    \caption{Maximum Half-chain entanglement entropy $S$ averaged over trajectories as a function of bond dimension $\chi$ in MPS simulation for MFIM with dephasing noise. We consider the system size $L = 100$ and compare the entanglement in (a) unraveling in the computational basis and (b) optimized unraveling for various dephasing rates $\gamma$. Dashed lines with empty circles and solid lines with filled circles represent data points in the area- and volume-law phases, respectively.}
    \label{fig:4}
\end{figure}

The optimized unraveling allows simulating the noisy dynamics of an arbitrarily large system based on MPS in an extended regime.
In the MPS simulation, the required bond dimension $\chi$ grows exponentially with the system size [$\chi \sim \mathcal{O}(\exp(L))$] in the volume-law phase, while the required $\chi$ is only a constant [$\chi \sim \mathcal{O}(1)$] in the area-law phase.
We use two unraveling schemes to simulate the noisy MFIM of system size $L = 100$, where the exact simulation is not possible.
Indeed, we find that, for a relatively small bond dimension $\chi \le 256$, the entanglement saturates, and the MPS can capture the trajectory dynamics for $\gamma \ge 0.2$ in the conventional unraveling scheme, whereas the entanglement saturates in an extended regime $\gamma \ge 0.1$ in the optimized basis (Fig.~\ref{fig:4}).

\textit{Discussion.---}
We have introduced optimized unraveling schemes for trajectory simulation of one-dimensional open system dynamics under arbitrary types of decoherence.
We focused on the single-qubit unraveling and showed that the classical simulation is efficient when the decoherence rate is above a significantly reduced but finite $p_c$ ($\gamma_c$).
These results open multiple directions for future research.

By unraveling multiple noisy channels at the same time, an efficient simulation may be possible for noisy dynamics with any non-vanishing decoherence rate $\gamma$.
The entanglement between the system and ancilla qudits is generated at the rate of $\gamma$.
Over a time period $T \gtrsim 1/\gamma$, each system qudit becomes decohered and only entangles with the ancilla qudits.
Thus, considering an optimized unraveling basis over $LT$ ancilla qudits for all noisy channels within this period, one can disentangle the system in each trajectory into a short-range entangled state~\cite{schumacher2002approximate}.
However, this procedure generally requires optimization over a nonlocal unitary rotation on $LT$ ancilla qudits, which is computationally challenging.
It remains open whether one can find an efficient representation of the optimized unraveling basis over $LT$ ancilla qudits.

Alternatively, one can take a coarse-grained perspective on noisy random circuits. 
Specifically, we combine all the gates and decoherence channels within a diamond-shaped region into a single quantum channel on two qudits, each involving $m$ consecutive qubits in the original circuit [as shown Fig.~\ref{fig:1} (a,c)].
Although each diamond-shaped block contains $\mathcal{O}(m^2)$ two-qubit unitary gates, only $\mathcal{O}(m)$ gates support on both qudits and generate entanglement between them.
In contrast, decoherence within each block generates $\mathcal{O}(m^2)$ entanglement between the system and ancilla qudits.
Thus, after the coarse-graining, the effective decoherence rate increases by $m$; we expect a reduced critical decoherence rate $\gamma_c = \calO(1/m)$ if we optimize the unraveling basis over the coarse-grained noisy channel.
Hence, by grouping $m \sim 1/ \gamma$ qubits in a single qudit and obtaining the optimized unraveling for noisy channels acting on two qudits, one can facilitate efficient simulation at an arbitrarily small but finite $\gamma$.
We leave for future work to develop an explicit algorithm to realize this idea.
We remark that such an algorithm only involves finding optimized unraveling for polynomially many decoherence channels that involve $\calO(m^2)$ ancilla qudits, which is at a cost of $\mathcal{O}(\exp(m^2)) \sim \mathcal{O}(\exp(1/\gamma^2))$. 
The polynomial resource in system size $L$ is consistent with previous results from both theoretical complexity analysis~\cite{aharonov2022polynomial} and numerical simulation based on matrix product operators~\cite{noh2020efficient}.

Another future direction is to study the optimized unraveling for open system dynamics in higher dimensions.
In this case, bipartite entanglement does not on its own determine the cost of classical simulation.
It remains open how to design an unraveling scheme to reduce the complexity of simulating trajectory dynamics on classical computers.

Furthermore, it is of great practical interest to simulate noisy evolution in near-term quantum simulation platforms based on the optimized trajectory unraveling.
On one hand, the classical simulation allows benchmarking near-term devices~\cite{choi2023preparing,mark2022benchmarking}.
On the other hand, the input from classical simulation combined with samples from quantum simulators may allow probing ``computationally assisted" observables~\cite{lee2022measurement,gullans2020scalable,garratt2023measurements} that are difficult to compute by classical computers alone.

\begin{acknowledgements}
\emph{Acknowledgement.---} The authors acknowledge Hannes
Pichler and Tianci Zhou for insightful discussions and
Matteo Ippoliti for helpful feedback on the manuscript. The
authors acknowledge support from the Challenge Institute
for Quantum Computation (NSF QLCI program, Grant
No. 2016245) and the Center for Ultracold Atoms (NSF
Physics Frontiers Center, Grant No. 1734011). Z. C. is
partially supported by the Defense Advanced Research
Projects Agency (DARPA) ONISQ program (Grant
No. 134371-5113608), Y. B. is partially supported by
NSF QLCI program (Grant No. OMA-2016245), and
S. C. is partially supported by NSF CAREER (Grant
No. 2237244).
\emph{Note added}: Upon completion of the present manuscript, we became aware of an independent work appearing on arXiv on the same day~\cite{Cheng_2023}, which also studies the optimized unraveling based on statistical mechanics mapping. However, the two works consider different research problems and have distinct motivations---this work focuses on unraveling in simulating one-dimensional open system dynamics whereas Ref.~\cite{Cheng_2023} focuses on unraveling in sampling from two-dimensional quantum circuits, which maps to simulating one-dimensional open system dynamics subject to measurements.
\end{acknowledgements}

\bibliography{refs}

\end{document}


\title{Supplementary Online Material for ``Optimized trajectory unraveling for classical simulation of noisy quantum dynamics''}

\author{Zhuo Chen}
\affiliation{Center for Theoretical Physics, Massachusetts Institute of Technology, Cambridge, MA 02139, USA}
\affiliation{The NSF AI Institute for Artificial Intelligence and Fundamental Interactions, Cambridge, MA 02139, USA}
\author{Yimu Bao}
\affiliation{Department of Physics, University of California, Berkeley, California 94720, USA}
\author{Soonwon Choi}
\affiliation{Center for Theoretical Physics, Massachusetts Institute of Technology, Cambridge, MA 02139, USA}

\maketitle

\tableofcontents

\section{Unraveling of quantum channels} \label{sec:unraveling}

General quantum channel $\calN$ can be decomposed into a set of $N$ Kraus operators, i.e.
\begin{align}
    \calN[\rho] = \sum_{\alpha = 0}^{N-1} K_\alpha \rho K_\alpha^\dagger,
\end{align}
where $K_\alpha$ satisfies $\sum_\alpha K^\dagger_\alpha K_\alpha = \id$.
In this work, we study three decoherence channels in qubit systems---dephasing, amplitude damping, and depolarization channels---as examples to demonstrate our optimized unraveling. 

\subsubsection{Dephasing channel}
The dephasing channel in a qubit system is defined as
\begin{equation}
    \calN_{\phi}[\rho] = \left(1-\frac{p}{2}\right)\rho + \frac{p}{2}Z\rho Z,
\end{equation}
where $Z$ is the Pauli-Z operator.

The conventional unraveling of the dephasing channel is based on projective measurements in a computational basis and can be represented using the following Kraus operators
\begin{equation}
\begin{aligned}
    K_0 = \sqrt{1-p}\id, \quad
    K_1 = \sqrt{p}\ketbra{0}, \quad
    K_2 = \sqrt{p}\ketbra{1}.
\end{aligned}
\end{equation}
Alternatively, the dephasing channel can be unraveled with a minimum number of two Kraus operators
\begin{equation} \label{eq:dephase_uni}
\begin{aligned}
    K_0 = \sqrt{1-\frac{p}{2}}\id, \quad
    K_1 = \sqrt{\frac{p}{2}}Z.
\end{aligned}
\end{equation}
In this unraveling, each Kraus operator is proportional to a unitary operator, and it has the interpretation of applying a $Z$ gate with probability $p/2$. 
We search for an optimized unraveling scheme related to this minimum set by a unitary transformation

\subsubsection{Amplitude damping channel}
Another example we consider is the amplitude damping channel
\begin{equation}
    \calN_{d}\left[\begin{pmatrix}
        \rho_{00} & \rho_{01} \\
        \rho_{10} & \rho_{11}
    \end{pmatrix} \right] = \begin{pmatrix}
        \rho_{00} + p\rho_{11} & \sqrt{1- p}\rho_{01} \\
        \sqrt{1 -p} \rho_{10} & (1 - p)\rho_{11}
    \end{pmatrix}
\end{equation}
where $p$ is the damping rate.
The channel describes spontaneous decay from $\ket{1}$ to $\ket{0}$ with a probability $p$. 

Conventionally, the channel is written in terms of two Kraus operators
\begin{equation} \label{eq:amp_damp}
\begin{aligned}
    K_0=\ketbra{0}{0} + \sqrt{1-p}\ketbra{1}{1}, \quad K_1=\sqrt{p}\ketbra{0}{1}.
\end{aligned}
\end{equation}
Here, we consider equivalent unraveling schemes that are related to the conventional unraveling by a unitary transformation.

\subsubsection{Depolarization channel}
The third example we consider is the depolarization channel that describes qubit loss at a probability $p$
\begin{equation}
    \calN_p[\rho] = \left(1-\frac{3p}{4}\right)\rho + \frac{p}{4}X\rho X + \frac{p}{4}Y\rho Y + \frac{p}{4}Z\rho Z.
\end{equation}

Conventionally, the depolarization channel is decomposed in terms of the Kraus operators
\begin{equation}
\begin{aligned}
    K_0=\sqrt{1-p}\id \quad 
    K_1=\sqrt{\frac{p}{2}}\ketbra{0}{0}, \quad
    K_2=\sqrt{\frac{p}{2}}\ketbra{0}{1}, \quad 
    K_3=\sqrt{\frac{p}{2}}\ketbra{1}{0}, \quad
    K_4=\sqrt{\frac{p}{2}}\ketbra{1}{1},
\end{aligned}
\end{equation}
which corresponds to replacing the noisy qubit with a fresh qubit, randomly sampled from $\{\ket{0}, \ket{1}\}$, with probability $p$. 
Alternatively, a minimum number of four Kraus operators is sufficient to describe the same depolarization channel, which reads
\begin{equation}
\begin{aligned}
    K_0 = \sqrt{1-\frac{3p}{4}}\id, \quad 
    K_1 = \sqrt{\frac{p}{4}}X, \quad 
    K_2 = \sqrt{\frac{p}{4}}Y, \quad 
    K_3 = \sqrt{\frac{p}{4}}Z.
\end{aligned}
\end{equation}
Similar to the dephasing channel, the optimal unraveling is parameterized as a unitary transformation from this minimum set.

\section{Optimized unraveling based on the effective spin model}

The quantum trajectories of noisy random circuits generally undergo monitored dynamics.
When tuning the decoherence rate, the trajectory averaged entanglement entropy can undergo a measurement-induced transition from a volume- to an area-law scaling~\cite{li2018quantum,skinner2019measurement,li2019measurement}, indicating a critical decoherence rate $p_c$ above which efficient classical simulation is possible.
Our goal is to find an optimized unraveling scheme that yields the lowest $p_c$.
In this section, we derive such an unraveling scheme for noisy random unitary circuits based on the effective spin model~\cite{bao2020theory,jian2020measurement}.
Here, we consider unraveling every decoherence channel in the same basis.
Moreover, we focus on the unraveling schemes with a minimum number of Kraus operators, which are related to the minimum set by a unitary transformation (see Sec.~\ref{sec:unraveling}).

The central quantity of interest is the trajectory averaged von Neumann entropy for subsystem $A$
\begin{align}
    S_A = \mathbb{E}_U\left[-\sum_m p_m \tr (\rho_{A,m} \log \rho_{A,m}) \right],
\end{align}
where $p_m$ is the probability for each quantum trajectory labeled by $m$, and $\mathbb{E}_U[\cdot]$ represents averaging over random circuit realizations.
However, analytically evaluating such a quantity is challenging.
Instead, one can consider a series of quasi-entropy that approximates the true von Neumann entropy
\begin{align}
    \tilde{S}^{(n)}_A = \frac{1}{1-n} \log\left( \frac{\mathbb{E}_U[\sum_m p_m^n \tr \rho_{A,m}^n]}{\mathbb{E}_U[\sum_m p_m^n]} \right).
\end{align}
The quasi-entropies recover the von Neumann entropy in the replica limit $n \to 1$, i.e.
\begin{align}
    S_A = \lim_{n \to 1} \tilde{S}^{(n)}_A.
\end{align}

These quasi-entropies admit an analytic understanding in terms of the excess free energy of domain walls in effective spin models~\cite{bao2020theory,jian2020measurement}.
These spin models undergo ferromagnetic transitions when tuning the decoherence rate, which manifest as the entanglement transition in the corresponding quasi-entropy.
Thus, by minimizing the critical decoherence rate associated with the ferromagnetic transition in the spin model, one can obtain an optimized unraveling basis.

In this work, we focus on the second quasi-entropy $n = 2$.
Although, the critical point $p_c^{(2)}$ associated with the second quasi-entropy only approximate $p_c$, the optimized unraveling scheme still significantly reduces the von Neumann entropy in trajectories as demonstrated by numerics in Sec.~\ref{suppsec:numerics}.
Remarkably, the corresponding spin model for $n = 2$ is exactly solvable, allowing us to analytically determine the unraveling scheme that gives the lowest critical decoherence rate.

Following the derivation in Ref.~\cite{bao2020theory}, the effective spin model for $n = 2$ is a two-dimensional classical Ising model on a triangular lattice as shown in Fig.~2(a) in the main text.
The model involves ferromagnetic $J_d$ and anti-ferromagnetic Ising coupling $J_h$ on the diagonal and horizontal bonds, respectively,
\begin{equation}
J_d = \frac{1}{4} \log\left(\frac{-u_2^2/q^2+u_1^2}{u_2^2-u_1^2/q^2}\right), \quad J_h = \frac{1}{4} \log\left\{\frac{[u_1u_2(1-1/q^2)]^2}{(u_2^2-u_1^2/q^2)(u_1^2-u_2^2/q^2)}\right\},
\end{equation}
where $q$ is the local Hilbert space dimension, and $u_1$ and $u_2$ are determined by the Kraus operators,
\begin{align}
    u_1 = \sum_{\alpha = 0}^{N-1} \left(\tr K_\alpha^\dagger K_\alpha K_\alpha^\dagger K_\alpha\right), \quad u_2 = \sum_{\alpha = 0}^{N-1} \left(\tr K_\alpha^\dagger K_\alpha\right)^2.
\end{align}
We note that \{$K_\alpha$\}, also both $u_1$ and $u_2$ are functions of $p$. 

The triangular lattice Ising model is exactly solvable~\cite{eggarter1975triangular,tanaka1978triangular}, and the critical point $p_c^{(2)}$ can be determined from 
\begin{equation}
    2e^{2J_h} = e^{-2J_d} - e^{2J_d},
\end{equation}
which is equivalent to
\begin{equation} \label{eq:critical_point}
    \left(\frac{u_2}{u_1}\right)^2-2\frac{q^2-1}{q^2+1}\left(\frac{u_2}{u_1}\right) - 1 = 0,
\end{equation}
For qubit system ($q=2$), the solution of Eq.~\eqref{eq:critical_point} reads
\begin{equation} \label{eq:critical_solution}
    \frac{u_2}{u_1} = \frac{1}{5}\left(3 + \sqrt{34}\right).
\end{equation}
One can then solve for the critical point $p_c^{(2)}$ given $u_1$ and $u_2$ as functions of $p$. 

The Kraus operators for different unraveling schemes are related by
\begin{equation}
    K'_\alpha = \sum_{\beta = 1}^{N-1} U_{\alpha \beta} K_\beta.
\end{equation}
While $U$ can be an isometry in general, we focus on decomposing the quantum channel in terms of a minimum number of Kraus operators and $U$ being a unitary rotation in this work.
Our goal is to find the unitary transformation that gives the minimum $p_c^{(2)}$.  
In general, it can be hard to solve for an analytical expression of $U$ that gives the minimum $p_c^{(2)}$; however, a numerical optimization algorithm is always possible, as long as the search space is small. 
For a qubit system, any local quantum channel can be represented with at most four Kraus operators. 
Therefore, we can always parameterize $U$ as using at most $15$ parameters (SU(4) group has 15 free parameters) and perform numerical optimization.
The spin-model optimized $p_c^{(2)}$ for dephasing, amplitude damping, and depolarization channels are summarized in Table~\ref{tab:pc_spin_model}.

\begin{table}[h!]
    \centering
    \begin{tabular}{|c|c|c|}
    \hline\hline
     Unraveling basis & Conventional & Optimized \\
     \hline
     Dephasing & 0.3558 & 0.0685 \\
     Depolarization & 0.4386 & 0.0457 \\
     Amplitude damping & 0.4205 & 0.1324 \\
    \hline\hline
    \end{tabular}
    \caption{Critical decoherence rate $p_c^{(2)}$ in the effective spin model for noisy random circuits. $p_c^{(2)}$ in the conventional and optimized unraveling scheme for dephasing, depolarization, and amplitude damping channel are presented.
    }
    \label{tab:pc_spin_model}
\end{table}

For dephasing and amplitude damping channels, the unraveling scheme that gives the minimum $p_c^{(2)}$ can be analytically determined.
Below, we derive the minimum $p_c^{(2)}$ and the associated Kraus operators for these two channels.

\subsubsection{Dephasing channel}
For the dephasing channel, we consider the Kraus operators related to the minimum set [Eq.~\eqref{eq:dephase_uni}] by a general SU(2) unitary transformation
\begin{align}
    \begin{pmatrix}
        K_0 \\
        K_1
    \end{pmatrix} = \begin{pmatrix}
        e^{\ri(\psi+\phi)}\cos \theta & \ri e^{\ri(\psi-\phi)}\sin \theta \\
        \ri e^{-\ri(\psi-\phi)}\sin \theta & e^{-\ri(\psi+\phi)}\cos \theta
    \end{pmatrix} \begin{pmatrix}
        \sqrt{1 - \frac{p}{2}}\id \\
        \sqrt{\frac{p}{2}}Z
    \end{pmatrix}.
\end{align}
Using this parameterization, the couplings are given by

\begin{align} \label{eq:dephase_u1}
    u_1 &= 2\left\{\left[\left(1-\frac{p}{2}\right)^2 + \left(\frac{p}{2}\right)^2\right]\left(\cos^4\!\theta+\sin^4\!\theta\right) + 4\left(1-\frac{p}{2}\right)\left(\frac{p}{2}\right)\cos^2\!\theta \sin^2\!\theta \left(2+\sin^2 2\phi\right) \right\}, \\
    u_2 &= 4\left\{\left[\left(1-\frac{p}{2}\right)^2 + \left(\frac{p}{2}\right)^2\right]\left(\cos^4\!\theta+\sin^4\!\theta\right)+4\left(1-\frac{p}{2}\right)\left(\frac{p}{2}\right)\cos^2\!\theta \sin^2\!\theta \right\},
\end{align}
which has no dependency on $\psi$.
Since $u_2/u_1$ is a non-increasing function of both $p$ and $\sin^2\!\phi$, there is a tradeoff between $p_c^{(2)}$ and $\sin^2 2\phi$ at the critical point [Eq.~\eqref{eq:critical_solution}]. Therefore, the minimum $p_c^{(2)}$ must be found when $\sin^2 2\phi$ is maximized ($\phi=\pi/4$ and $\sin^2 2\phi = 1$). 
Then, one can search along $\theta$ to find the optimal $p_c^{(2)}=0.0685$ at $\theta=\pi/4$. Choosing $\psi=-\pi/4$ (real-valued unitary transformation), the resulting Kraus operators are
\begin{equation}
    \begin{pmatrix}
        K_0 \\
        K_1
    \end{pmatrix} = \frac{1}{\sqrt{2}}\begin{pmatrix}
        1 & 1 \\
        -1 & 1
    \end{pmatrix} \begin{pmatrix}
        \sqrt{1 - \frac{p}{2}}\id \\
        \sqrt{\frac{p}{2}}Z
    \end{pmatrix}.
\end{equation}

\subsubsection{Amplitude damping channel}
The Kraus operator for the amplitude damping channel can be derived analogously. We consider a SU(2) unitary transformation of the conventional unraveling [Eq.~\eqref{eq:amp_damp}] as
\begin{align}
    \begin{pmatrix}
        K_0 \\
        K_1
    \end{pmatrix} = \begin{pmatrix}
        e^{\ri(\psi+\phi)}\cos \theta & \ri e^{\ri(\psi-\phi)}\sin \theta \\
        \ri e^{-\ri(\psi-\phi)}\sin \theta & e^{-\ri(\psi+\phi)}\cos \theta
    \end{pmatrix} \begin{pmatrix}
        \ketbra{0}{0} + \sqrt{1-p}\ketbra{1}{1} \\
        \sqrt{p}\ketbra{0}{1}
    \end{pmatrix}.
\end{align}
This gives rise to
\begin{align}
    u_1 &= 2(1-p+p^2)\left(\cos^4\!\theta+\sin^4\!\theta\right)  + 4p(2-p)\cos^2\!\theta \sin^2\!\theta, \\
    u_2 &= 2(2-2p+p^2)\left(\cos^4\!\theta+\sin^4\!\theta\right)  + 4p(2-p)\cos^2\!\theta \sin^2\!\theta.
\end{align}
Here, $u_1$ and $u_2$ are only functions of $\theta$, where
the unraveling that minimizes $p_c^{(2)}$ can be found at $p_c^{(2)}=0.1324$ compared to the conventional unraveling with $p_c^{(2)}=0.4205$. Choosing $\psi=-\pi/4$ and $\phi=\pi/4$ (real-valued unitary transformation), the resulting Kraus operators are
\begin{equation}
    \begin{pmatrix}
        K_0 \\
        K_1
    \end{pmatrix} = \frac{1}{\sqrt{2}}\begin{pmatrix}
        1 & 1 \\
        -1 & 1
    \end{pmatrix} \begin{pmatrix}
        \ketbra{0}{0} + \sqrt{1-p}\ketbra{1}{1} \\
        \sqrt{p}\ketbra{0}{1}
    \end{pmatrix}.
\end{equation}

\section{Details for heuristic optimization algorithm}
For the heuristic optimization method, we minimize the average entropy of the noisy system qubit averaged over the measurement outcomes of the ancilla qudit. 
We note that since the entropy of the noisy qubit is state-dependent, the optimized unraveling depends on previous measurement outcomes. 
In this sense, the optimized unraveling scheme in the heuristic algorithm is, strictly speaking, achieved by a nonlocal unitary transformation on ancilla qudits, which involves local operations and classical communication.
In other words, this algorithm is adaptive.
Moreover, although decoherence channels at the same time step commute with each other, the order in which their unraveling basis is optimized could potentially affect the resulting state in trajectories. 
In this work, we choose the order of the optimization sequentially from qubit $1$ to qubit $L$ for dephasing channels within a single time step.

\section{Trotterization of Lindblad equation}
For a generic Lindblad equation
\begin{equation}
    \dot \rho = -i \comm{H}{\rho} + \sum_i\gamma_i \left(L_i \rho L_i^\dagger - \frac{1}{2}\acomm{L_i^\dagger L_i}{\rho}\right),
\end{equation}
we can trotterize the unitary evolution with respect to $H$ and the quantum jumps with respect to $\{L_i\}$, as
\begin{equation}
    \rho(t+dt) = \calN_{dt}\left[e^{-\ri H dt} \rho(t)e^{\ri H dt}\right] + \calO(dt^2),
\end{equation}
where the decoherence channel $\calN_{dt}$ is further decomposed into product of local channels, i.e. $\calN_{dt} = \prod_i \calN_{i, dt}$.
Each local channel only involves the jump operators $L_i$ that apply on a single qubit and is described by the following Kraus operators
\begin{equation}
    K_0 = \id - \frac{dt}{2}\sum_{i}\gamma_i L_i^\dagger L_i, \quad K_{\alpha > 0} = \sqrt{\gamma_\alpha dt} L_\alpha.
\end{equation}
The Hamiltonian evolution operator $e^{-\ri H dt}$ is also trotterized to achieve the brick-layer structure as used in this work. 

For dephasing error where $L_i = Z_i$ and $\gamma_i=\gamma$, we arrive at 
\begin{equation}
    K_0 = \left(1-\frac{dt}{2}\gamma\right)\id, \quad K_{1} = \sqrt{\gamma dt}Z_i,
\end{equation}
which agrees with the unital unraveling of the dephasing channel with $p=2\gamma dt$ up to the lowest order in $dt$.

\section{Additional numerical results}\label{suppsec:numerics}
In this section, we provide additional numerical results for the critical decoherence rates associated with the von Neumann entropy and Renyi entropies in three different unravelings of the dephasing channel. 
The extracted critical decoherence rate $p_c$ for RUC and MFIM are presented in Table~\ref{tab:pc_ruc} and~\ref{tab:pc_mfim}, respectively.
The tripartite mutual information as a function of decoherence rate $p$ and the finite-size scaling analysis used to determine $p_c$ can be found in Fig.~\ref{fig:ruc_comp}-\ref{fig:mfim_optim}

\begin{table}[h!]
    \centering
    \begin{tabular}{|c|c|c|c|}
    \hline\hline
     Unraveling basis & Conventional & Spin-model Optimized & Heuristically Optimized \\
     \hline
     Renyi-$1/2$ & 0.196(3)   & 0.097(3)  & 0.096(3)\\
     Von Neumann & 0.168(3)  &  0.089(3) & 0.085(2)\\
     Renyi-$2$ & 0.165(4)  &  0.102(3) & 0.102(3)\\
     Renyi-$\infty$ & 0.172(5)  & 0.110(4)  & 0.110(3)\\
    \hline\hline
    \end{tabular}
    \caption{Critical decoherence rate $p_c$ for RUC with dephasing noise for various Renyi indices extracted from finite size scaling collapse.
    }
    \label{tab:pc_ruc}
\end{table}

\begin{table}[h!]
    \centering
    \begin{tabular}{|c|c|c|c|}
    \hline\hline
     Unraveling basis & Conventional & Spin-model Optimized & Heuristically Optimized \\
     \hline
     Renyi-$1/2$ & 0.222(5)  & 0.121(7)  & 0.116(8)\\
     Von Neumann & 0.201(6)  &  0.111(6) & 0.100(6)\\
     Renyi-$2$ & 0.196(7)   &  0.126(6) & 0.119(9)\\
     Renyi-$\infty$ & 0.205(8)  &  0.135(8) & 0.124(8)\\
    \hline\hline
    \end{tabular}
    \caption{Critical decoherence rate $\gamma_c$ for MFIM with dephasing noise for various Renyi indices extracted from finite size scaling collapse.
    }
    \label{tab:pc_mfim}
\end{table}

\begin{figure}
    \centering
    \includegraphics[width=0.5\linewidth]{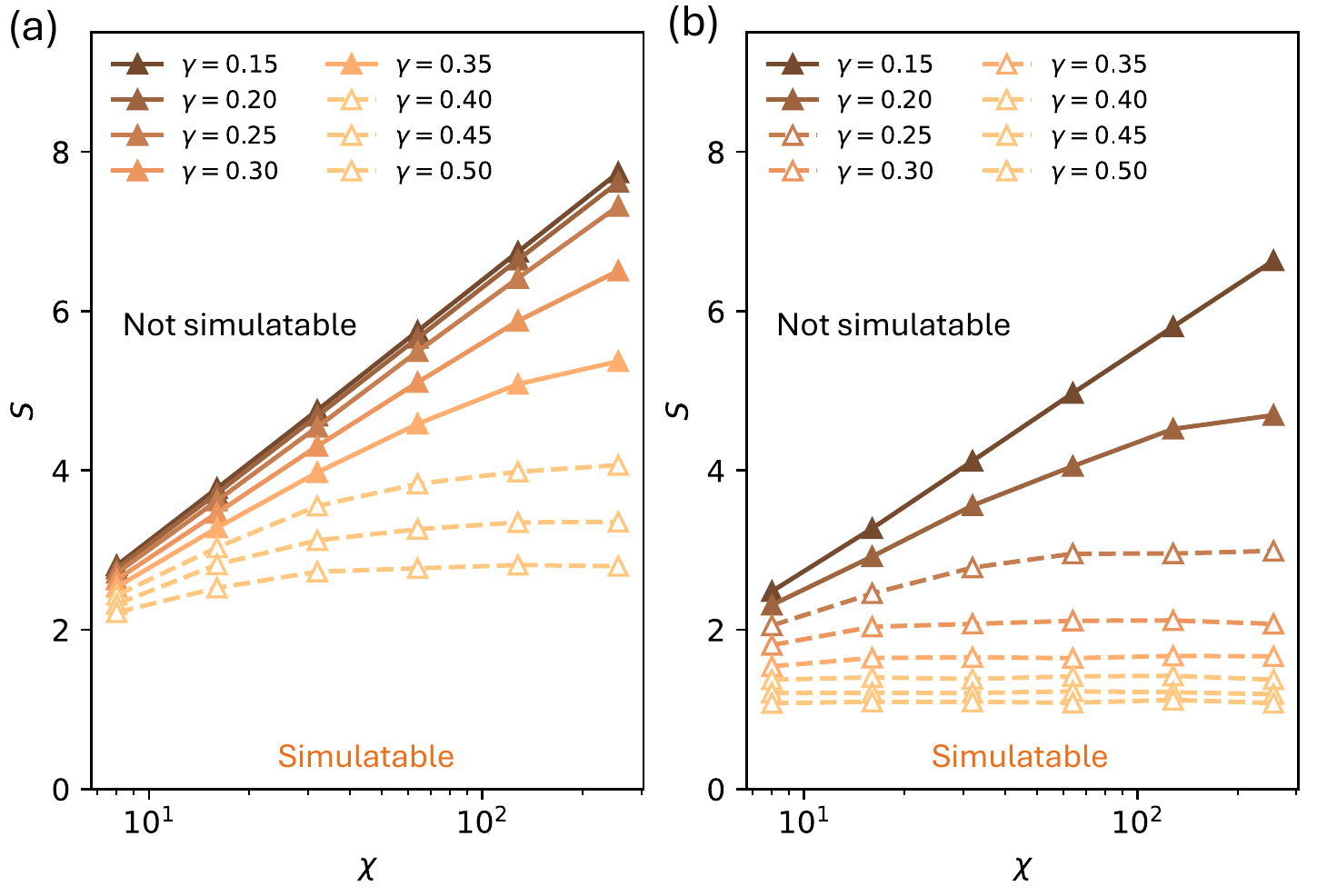}
    \caption{Maximum Half-chain entanglement entropy $S$ averaged over trajectories as a function of bond dimension $\chi$ in MPS simulation for MFIM with amplitude damping noise. We consider the system size $L = 100$ and compare the entanglement in (a) unraveling in the computational basis and (b) optimized unraveling for various noise rates $\gamma$. Dashed lines with empty circles and solid lines with filled circles represent data points in the area- and volume-law phases, respectively.}
    \label{fig:amp_damp}
\end{figure}

\begin{figure}[H]
    \centering
    \includegraphics[width=0.9\linewidth]{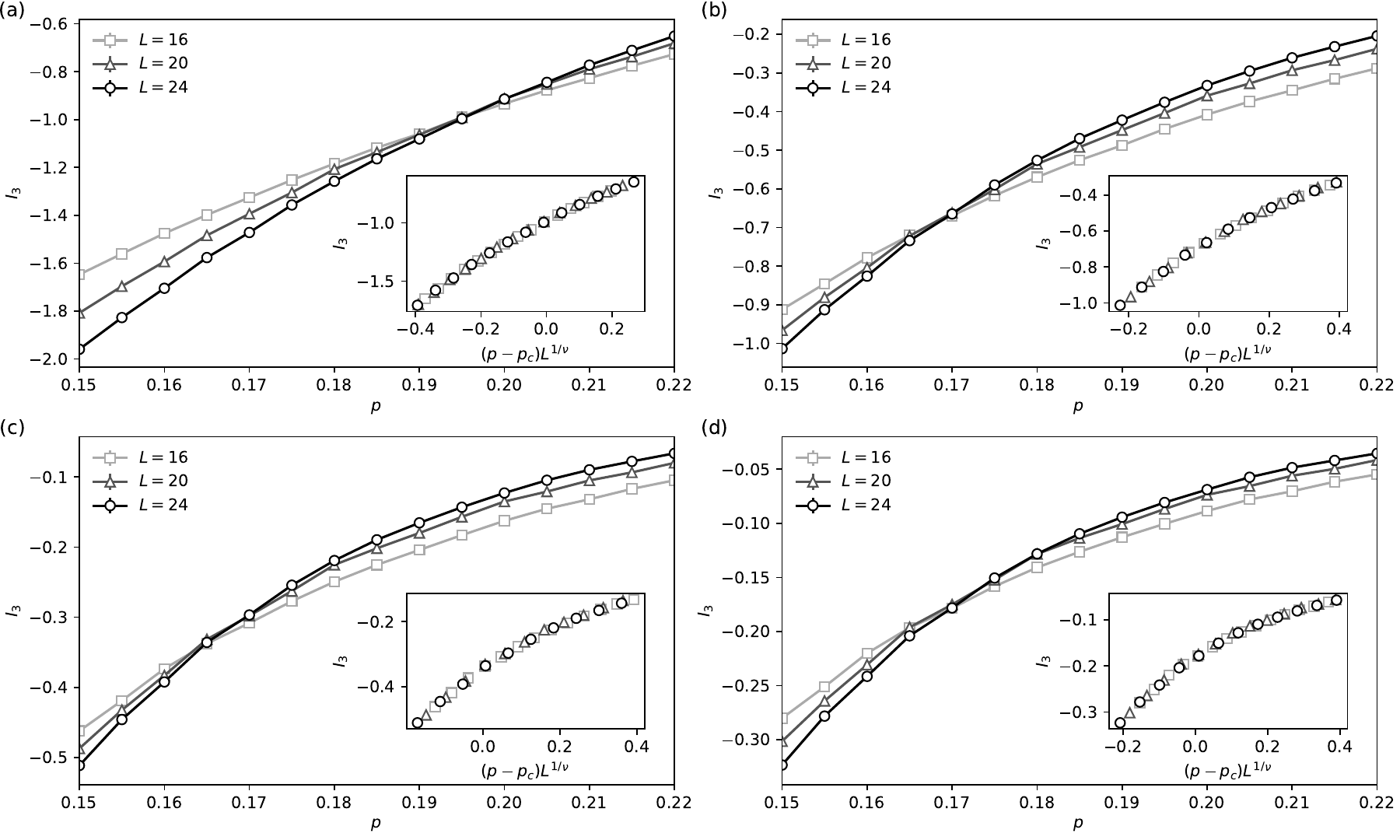}
    \caption{Tripartitie mutual information $I_3$ as a function of the decoherence rate $p$ in the computational unraveling basis for RUC. (Inset) Finite-size scaling collapse to determine the critical decoherence rate. Subplots: (a) Renyi-1/2 entropy; (b) von Neumann entropy; (c) Renyi-2 entropy; and (d) Renyi-$\infty$ entropy. The results are averaged over 400 quantum trajectories. The results here are consistent with Ref.~\cite{zabalo2020critical}.
    }
    \label{fig:ruc_comp}
\end{figure}

\begin{figure}[H]
    \centering
    \includegraphics[width=0.9\linewidth]{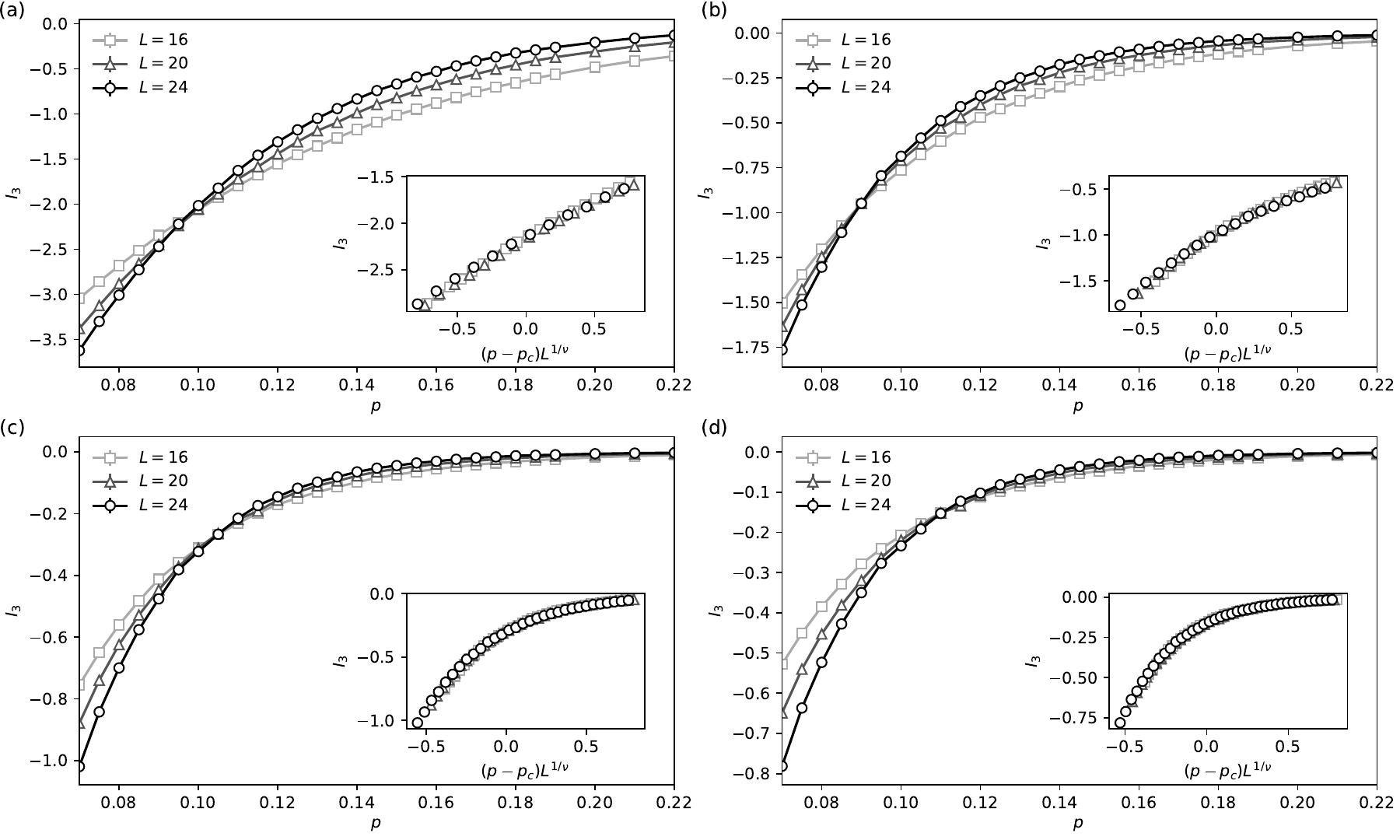}
    \caption{Tripartitie mutual information $I_3$ as a function of the decoherence rate $p$ in the spin-model optimized unraveling basis for RUC. (Inset) Finite-size scaling collapse to determine the critical decoherence rate. Subplots: (a) Renyi-1/2 entropy; (b) von Neumann entropy; (c) Renyi-2 entropy; and (d) Renyi-$\infty$ entropy. The results are averaged over 400 quantum trajectories.
    }
    \label{fig:ruc_spin}
\end{figure}

\begin{figure}[H]
    \centering
    \includegraphics[width=0.9\linewidth]{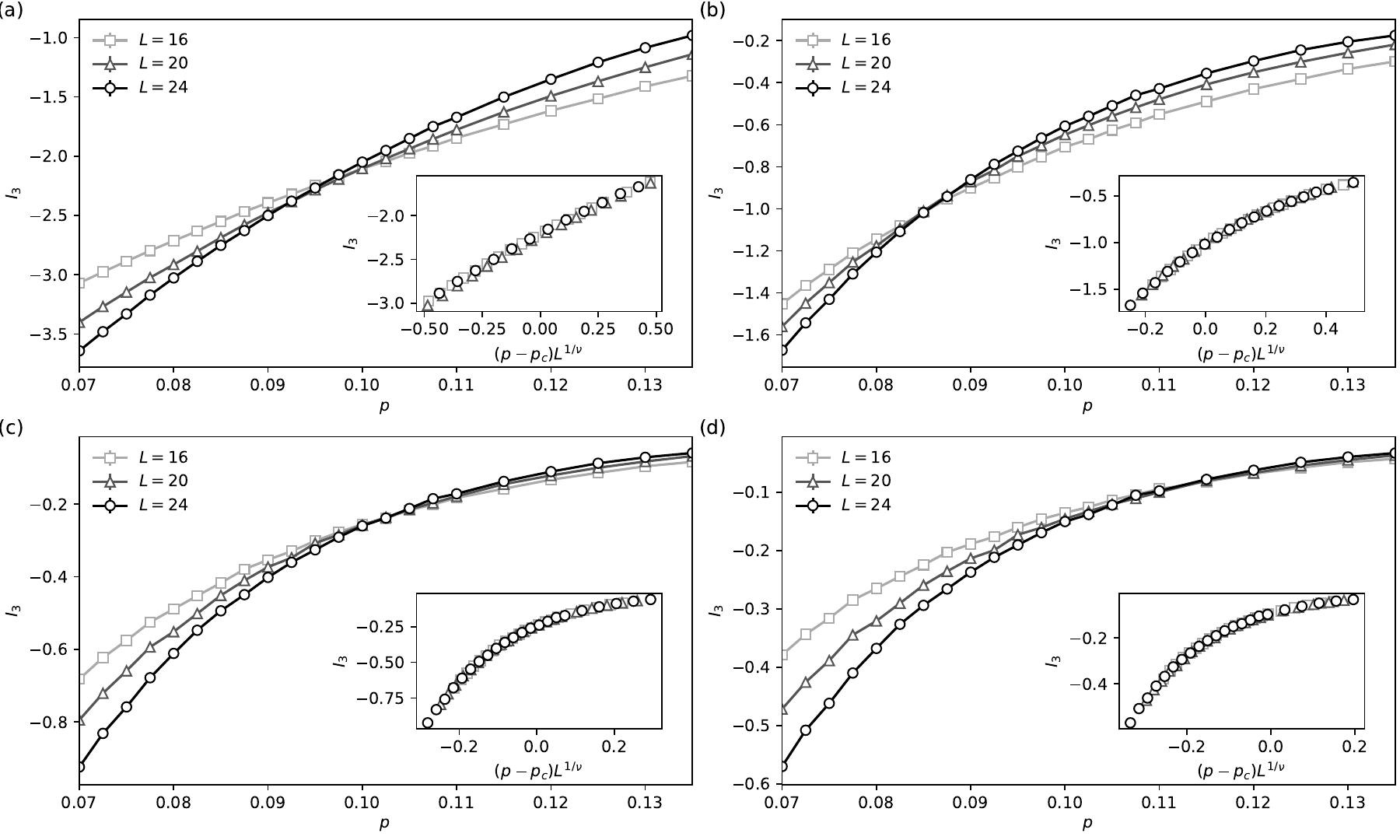}
    \caption{Tripartitie mutual information $I_3$ as a function of the decoherence rate $p$ in the heuristically optimized unraveling basis for RUC. (Inset) Finite-size scaling collapse to determine the critical decoherence rate. Subplots: (a) Renyi-1/2 entropy; (b) von Neumann entropy; (c) Renyi-2 entropy; and (d) Renyi-$\infty$ entropy. The results are averaged over 400 quantum trajectories.
    }
    \label{fig:ruc_optim}
\end{figure}

\begin{figure}[H]
    \centering
    \includegraphics[width=0.9\linewidth]{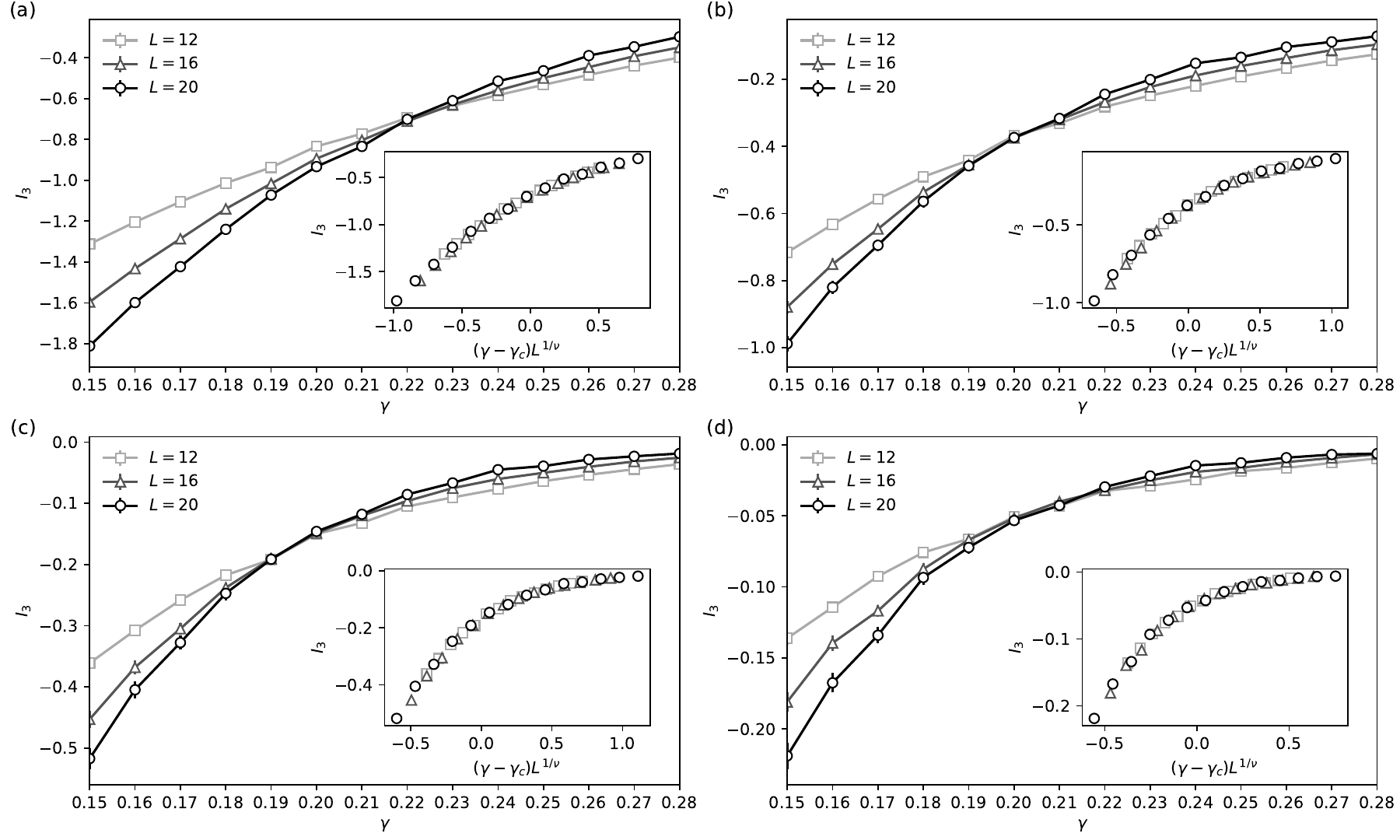}
    \caption{Tripartitie mutual information $I_3$ as a function of the decoherence rate $\gamma$ in the computational unraveling basis for MFIM. (Inset) Finite-size scaling collapse to determine the critical decoherence rate. Subplots: (a) Renyi-1/2 entropy; (b) von Neumann entropy; (c) Renyi-2 entropy; and (d) Renyi-$\infty$ entropy. The results are averaged over 100 quantum trajectories. 
    }
    \label{fig:mfim_comp}
\end{figure}

\begin{figure}[H]
    \centering
    \includegraphics[width=0.9\linewidth]{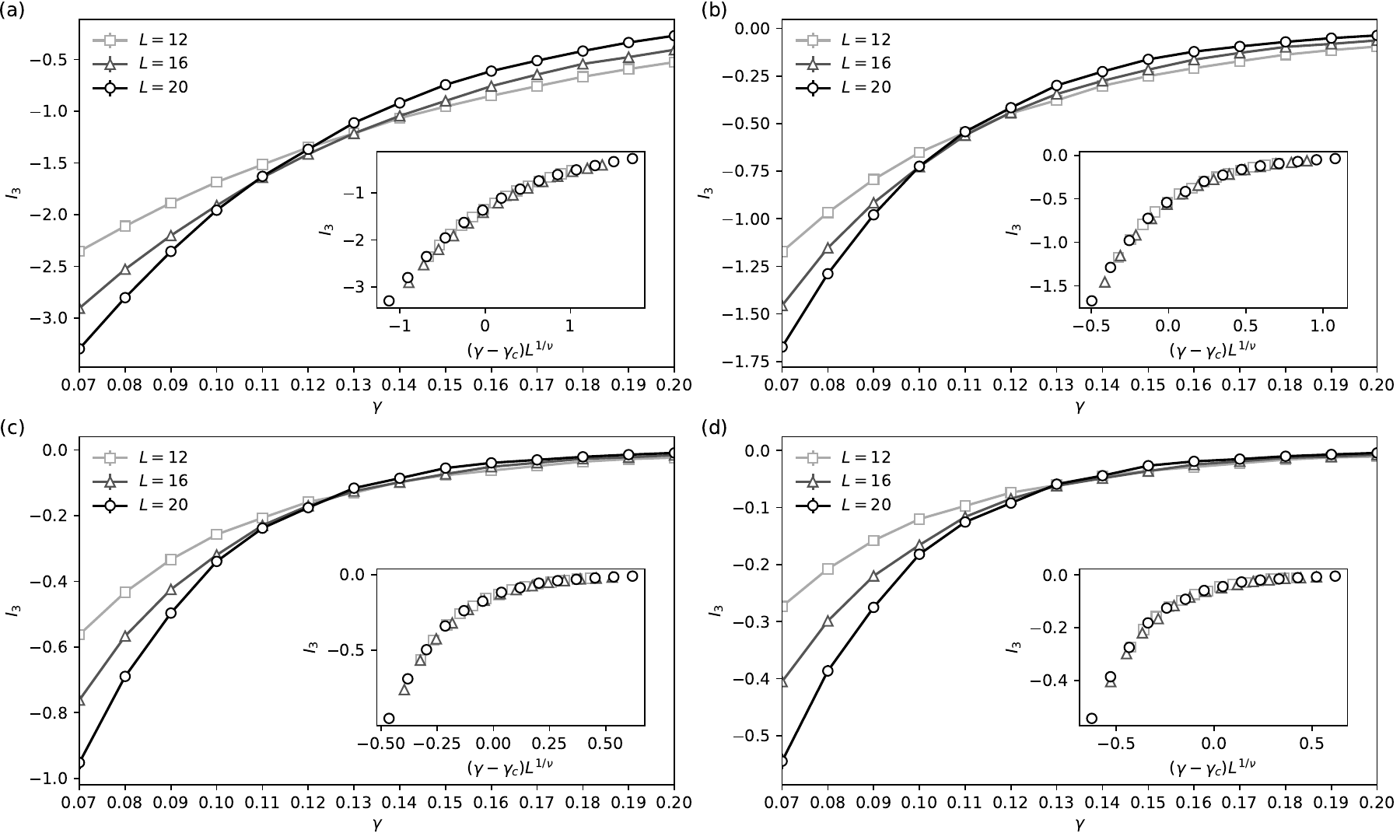}
    \caption{Tripartitie mutual information $I_3$ as a function of the decoherence rate $\gamma$ in the spin-model optimized unraveling basis for MFIM. (Inset) Finite-size scaling collapse to determine the critical decoherence rate. Subplots: (a) Renyi-1/2 entropy; (b) von Neumann entropy; (c) Renyi-2 entropy; and (d) Renyi-$\infty$ entropy. The results are averaged over 100 quantum trajectories. 
    }
    \label{fig:mfim_spin}
\end{figure}

\begin{figure}[H]
    \centering
    \includegraphics[width=0.9\linewidth]{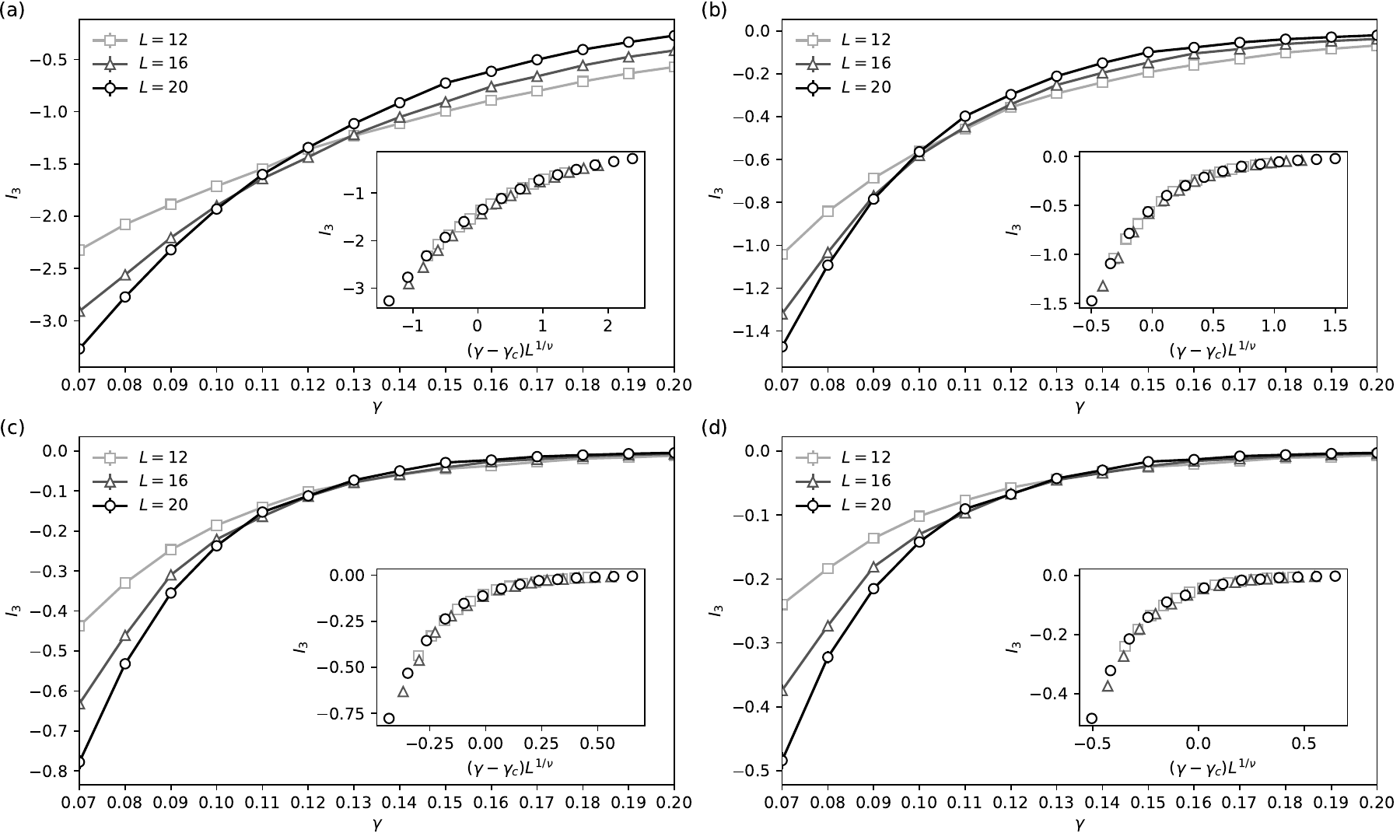}
    \caption{Tripartitie mutual information $I_3$ as a function of the decoherence rate $\gamma$ in the heuristically optimized unraveling basis for MFIM. (Inset) Finite-size scaling collapse to determine the critical decoherence rate. Subplots: (a) Renyi-1/2 entropy; (b) von Neumann entropy; (c) Renyi-2 entropy; and (d) Renyi-$\infty$ entropy. The results are averaged over 100 quantum trajectories. 
    }
    \label{fig:mfim_optim}
\end{figure}

In addition, we repeat the experiment of optimized unraveling scheme on the mixed-field Ising model (MFIM) with amplitude sampling noise in the Y direction
\begin{equation}
    \dot \rho = -\ri \comm{H}{\rho} + \gamma \sum_i  (L_i \rho L_i^\dagger - \frac{1}{2}\{L_i^\dagger L_i,\rho\})
\end{equation}
where $L_i = (Z_i - \ri X_i) / 2$. The result is shown in Fig.~\ref{fig:amp_damp}. As shown in the figure, for a relatively small bond dimension $\chi \le 256$, the entanglement saturates, and the MPS can capture the trajectory dynamics for $\gamma \ge 0.4$ in the conventional unraveling scheme, whereas the entanglement saturates in an extended regime $\gamma \ge 0.25$ in the optimized basis. This is consistent with both the dephasing noise result and Table~\ref{tab:pc_spin_model}.



\section{Comparison between optimized unraveling algorithm and MPO-based algorithm}

In this section, we compare our optimized unraveling algorithm implemented using MPS with the MPO-based algorithm on the mixed-field Ising model 
 (MFIM) model with dephasing noise.

\begin{figure}[ht!]
    \centering
    \includegraphics[width=0.6\linewidth]{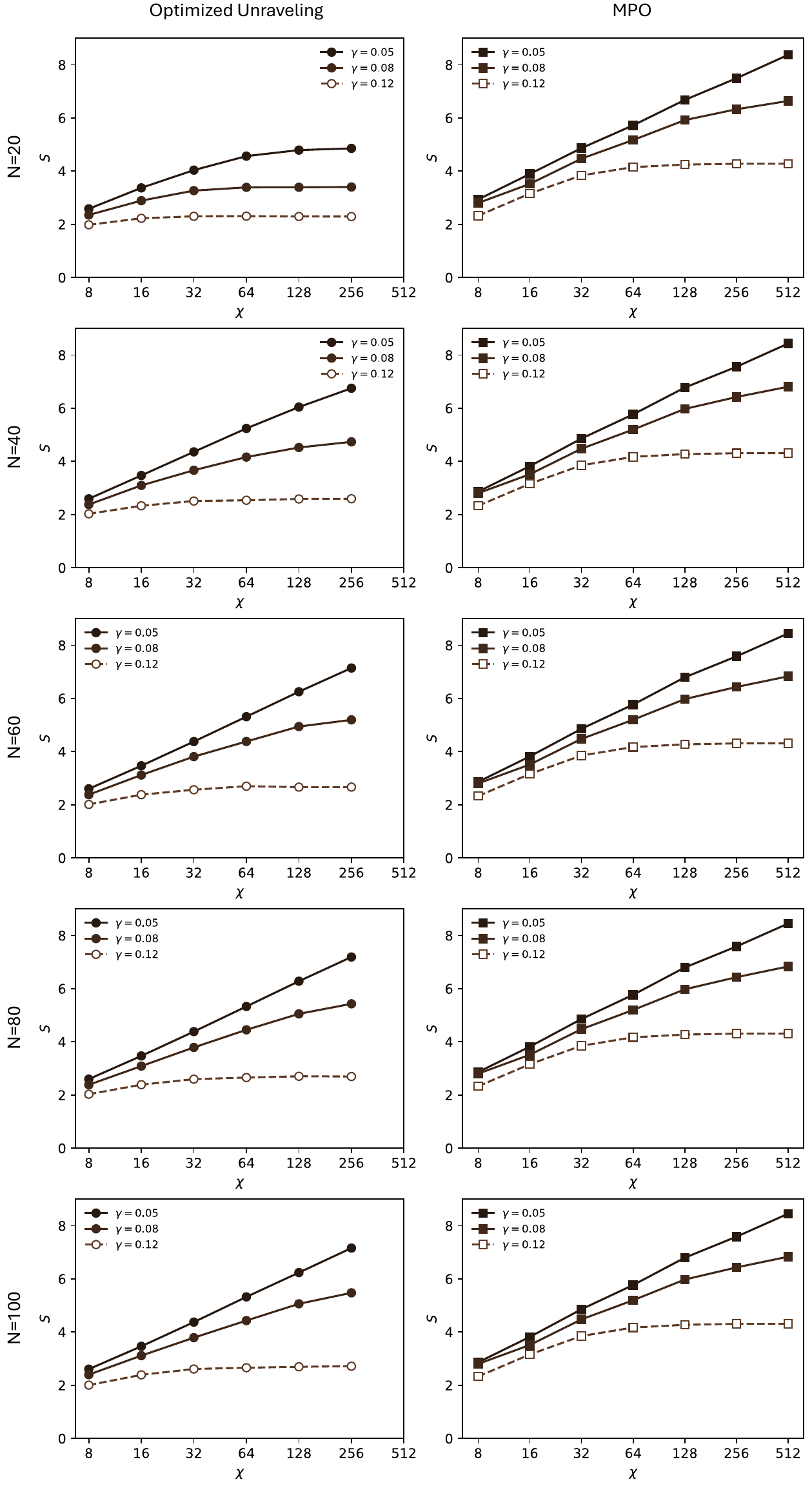}
    \caption{Maximum Half-chain entanglement entropy averaged over trajectories for the optimized unraveling method implemented on MPS, and the Maximum MPO entanglement entropy for the MPO-based method as a function of bond dimension for MFIM with dephasing noise. We consider different system sizes and compare the entanglement for various dephasing rates $\gamma$. Dashed lines with empty markers and solid lines with filled markers represent data points in the area- and volume-law phases for our method, respectively.
    }
    \label{fig:mps_vs_mpo}
\end{figure}

As mentioned in the main text, the MPO-based algorithm has issues with maintaining positivity and controlling errors. In contrast, our method does not encounter these issues despite requiring additional sampling; the ensemble of wave functions is guaranteed to be positive and has a controlled error. For small noise rates, e.g., $\gamma = 0.08$ (volume-law phase in our method), although MPO simulation is area-law in theory, the computational cost remains formidable because the required bond dimension is exponential in $1/\gamma$ (see Ref.~\cite{noh2020efficient}). Furthermore, in the limit of small noise rates and finite system sizes,the MPO-based algorithm required a bond dimension that is quadratic compared to that of MPS-based methods.

To illustrate this point, in Fig.~\ref{fig:mps_vs_mpo}, we run the MPO simulation for the same MFIM and plot the MPO entanglement entropy $S$ as a function of bond dimension $\chi$ compared to the MPS entanglement entropy. As shown in the figure, for large system sizes, the MPO entanglement entropy has not saturated at $\chi=512$ for $\gamma=0.05$ and $\gamma=0.08$, despite having an area law in theory. Therefore, in practice, $\gamma = 0.08$ is not simulatable for both the MPO method and our method in this regime. However, for small system size $N=20$, the entanglement entropy of our method saturates at $\chi=128$ even for small error rates $\gamma=0.05$, whereas MPO entanglement entropy does not saturate. This regime is important for simulating near-term quantum devices \cite{choi2023preparing,shaw_2024,shaw2024universalfluctuationsnoiselearning}; therefore, our method can bring additional practical advantages even in the volume law phase. 

For an intermediate error rate, $\gamma = 0.12$, when both methods seem manageable, the MPO method can still have a higher computational cost. Specifically, the entanglement entropy for the MPO method saturates at around $\chi = 64$, whereas for our method it saturates earlier at around $\chi=32$. Additionally, the MPO method has a larger local dimension ($d=4$ for spin-$1/2$ particles). For a TEBD-like algorithm, the simulation cost is $O(d^3\chi^3 + d^4\chi^2)$. The larger bond dimension and local dimension mean that the MPO simulation is at least 64 times more costly than an MPS simulation. Moreover, an even larger bond dimension may be required for MPO to compensate for the issues with unphysical solutions and uncontrolled errors as mentioned previously, further increasing the simulation cost.

In addition, although our method requires sampling wave function trajectories, such a process can be easily parallelized, with sampling errors well controlled in terms of the $1/\sqrt{M}$ scaling, with $M$ being the number of trajectories. Also, the convergence of observables of interest is often achieved significantly faster than the convergence of states, as in the case of many classical systems. Therefore, our method can be more efficient than the MPO method in practice, with the added benefit of maintaining physical solutions.

\bibliography{refs}